# Scalable parallel physical random number generator based on a superluminescent LED

Xiaowen Li,[1,2,4,*] Adam B. Cohen,[2,3] Thomas E. Murphy,[2,5] and Rajarshi Roy[2,3,4]

[1]*Department of Physics, Beijing Normal University, Beijing 100875, China*
[2]*Institute for Research in Electronics and Applied Physics, The University of Maryland, College Park, Maryland 20742, USA*
[3]*Department of Physics, The University of Maryland, College Park, Maryland 20742, USA*
[4]*Institute for Physical Science and Technology, The University of Maryland, College Park, Maryland 20742, USA*
[5]*Department of Electrical and Computer Engineering, The University of Maryland, College Park, Maryland 20742, USA*
*Corresponding author: xwli@bnu.edu.cn

We describe an optoelectronic system for simultaneously generating parallel, independent streams of random bits using spectrally separated noise signals obtained from a single optical source. Using a pair of nonoverlapping spectral filters and a fiber-coupled superluminescent LED (SLED), we produced two independent 10 Gb/s random bit streams, for a cumulative generation rate of 20 Gb/s. The system relies principally on chip-based optoelectronic components that could be integrated in a compact, economical package. © 2011 Optical Society of America

Random number generators are widely used in applications ranging from commerce to science, including secure key generation, communication, gaming, and Monte Carlo simulation. Many of the fastest physical random number generators reported to date have employed optoelectronic techniques, such as photon counting, homodyne detection of vacuum fluctuations, chaotic lasers, laser phase noise, and spontaneous emission [1–9]. While spontaneous emission has been recently exploited for random number generation [9], this Letter describes the first demonstration, to our knowledge, that multiple independent streams can be simultaneously obtained from a single optoelectronic noise source without the need for external optical amplification or gain. This represents an important step toward a chip-based ultrafast parallel physical random number generator.

Parallel random bit generation methods could dramatically improve the generation rate and scalability by producing many random bits simultaneously. Parallel pseudorandom number generation algorithms have been developed [10] and parallel hardware pseudorandom number generation methods based on cellular automata have been demonstrated [11–13]. With the exception of spatially multiplexed systems based on two-dimensional imaging [14,15], high-speed optical random number generators produce serial binary sequences that cannot be easily scaled to higher rates. Here, we describe a physical random number generator that utilizes wavelength demultiplexing filters to produce multiple, independent, high-speed random bitstreams from a single broadband optical source. Specifically, we demonstrate simultaneous generation of two statistically independent 10 Gb/s random bit streams using spectrally distinct signals obtained from a single superluminescent LED (SLED). We quantify the independence of the two streams by statistically computing the cross correlation and mutual information between them, and we test the quality of each stream using statistical tests developed for cryptographic security.

In Fig. 1, we show the experimental setup of the multichannel parallel random number generator. A fiber-coupled SLED (Exalos 1520) generates broadband optical noise that is separated into two nonoverlapping spectral slices using a pair of optical filters. Figure 2(a) plots the measured power spectrum of the amplified spontaneous emission (ASE) generated by the SLED, together with the spectrum transmitted by the two wavelength-division multiplexing filters. The filters each had an optical transmission bandwidth of $2.2\,\text{nm}$, and center wavelengths of 1540 and $1555\,\text{nm}$, respectively. The two filtered signals are detected by a pair of $11\,\text{GHz}$ photoreceivers (Discovery DSC-R402). Figure 2(b) plots the measured electrical spectrum obtained from one photoreceiver, in comparison to the background electrical noise obtained by extinguishing the SLED. The spectrally sliced ASE produces a fast, fluctuating electrical signal that is much stronger than the background electrical noise.

Each of the two received analog signals is threshold detected by a clocked comparator to produce a sequence of random bits that are recorded and processed using a pair of bit error rate testers (BERTs). An external 10 GHz clock signal determines the bit generation rate, and a second external trigger signal is used to initiate the data acquisition process, ensuring that both BERTs acquire simultaneous and synchronous records. Bias tees are used to adjust the logic decision thresholds with a precision of 0.1 mV to balance the proportion of ones and zeros in each channel.

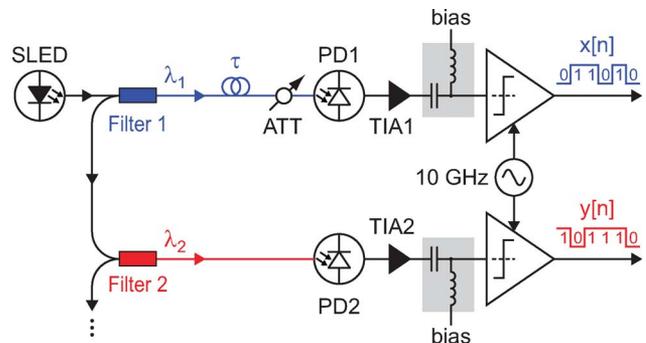

Fig. 1. (Color online) Optoelectronic system used to generate parallel random bit sequences. A SLED is spectrally sliced into multiple channels, each of which is detected with a $11\,\text{GHz}$ photodiode and transimpedance amplifier. A variable delay ($\tau$) and attenuator are used to equalize the time delay and power between the channels. The resulting signals are threshold detected using clocked comparators to simultaneously produce parallel independent binary sequences $x[n]$ and $y[n]$. PD, photodiode; TIA, transimpedance amplifier; ATT, attenuator.

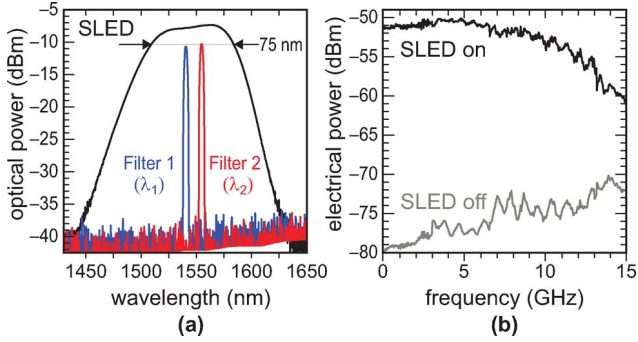

Fig. 2. (Color online) (a) Optical spectra of the SLED and the two spectrally sliced channels measured with a resolution bandwidth of 0.1 nm. (b) Electrical spectrum of the optical noise measured directly from one photoreceiver and the electrical background noise obtained by extinguishing the optical input signal. Both spectra were measured with a resolution bandwidth of 3 MHz.

In Fig. 3(a), we show representative bitmap images constructed from 1024 bits of data from the two channels of the system. The third panel shows the computed exclusive or (XOR) between the two binary sequences, which, for independent unbiased sequences, should exhibit no apparent pattern, bias, or correlation with the original sequences. To better quantify the independence of the two sequences, we statistically computed the correlation $\rho$ between the two binary sequences $x[n]$ and $y[n+k]$ as a function of their relative delay $k$:

$$\rho_{xy}[k] = \frac{\langle xy \rangle - \langle x \rangle \langle y \rangle}{\sqrt{(\langle x^2 \rangle - \langle x \rangle^2)(\langle y^2 \rangle - \langle y \rangle^2)}}, \quad (1)$$

where $\langle x \rangle$ and $\langle y \rangle$ denote the proportion of ones in $x[n]$ and $y[n+k]$, respectively, and $\langle xy \rangle$ indicates the proportion of ones of the product sequence $x[n]y[n+k]$. A related quantity is the mutual information [16], $I_{xy}$, which, for binary sequences, can be calculated as

$$I_{xy}[k] = \langle \bar{x}\bar{y} \rangle \log_2\left(\frac{\langle \bar{x}\bar{y} \rangle}{\langle \bar{x} \rangle \langle \bar{y} \rangle}\right) + \langle \bar{x}y \rangle \log_2\left(\frac{\langle \bar{x}y \rangle}{\langle \bar{x} \rangle \langle y \rangle}\right)$$
$$+ \langle x\bar{y} \rangle \log_2\left(\frac{\langle x\bar{y} \rangle}{\langle x \rangle \langle \bar{y} \rangle}\right) + \langle xy \rangle \log_2\left(\frac{\langle xy \rangle}{\langle x \rangle \langle y \rangle}\right), \quad (2)$$

where $\bar{x}$ indicates the logical complement of $x$, and $\langle \bullet \rangle$ again denotes the statistically evaluated proportion of ones. For binary sequences with low correlation and bias, the mutual information and correlation can be shown to be related by $I_{xy} \simeq \rho_{xy}^2/(2\ln 2)$. In Fig. 3(b), we plot the calculated correlation magnitude $|\rho_{xy}|$ and mutual information $I_{xy}$ as a function of delay $k$, for two $10^9$ bit sequences. The solid line shows the theoretical median level one would expect for two independent, unbiased $10^9$ bit sequences.

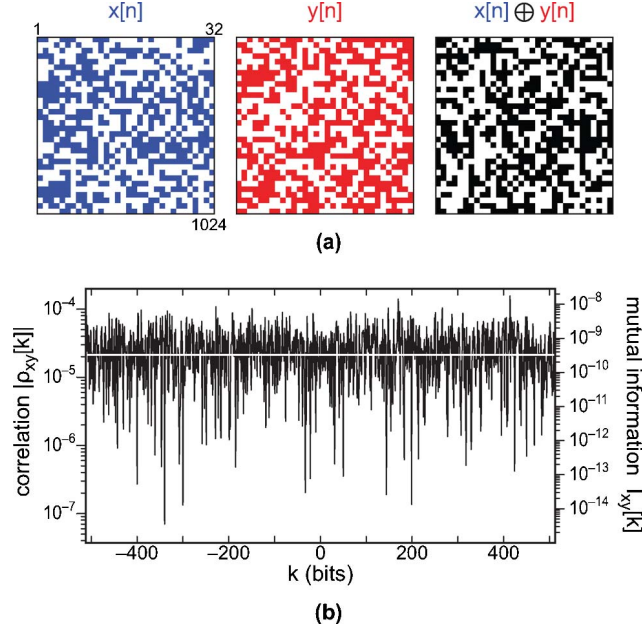

Fig. 3. (Color online) (a) Representative bitmap images constructed from the two data channels $x[n]$ and $y[n]$. The third panel shows the XOR between the two channels, $x[n] \oplus y[n]$, which exhibits no apparent bias, pattern, or correlation with the original sequences. (b) Statistically computed absolute correlation $|\rho_{xy}|$ and mutual information $I_{xy}[k]$ between the two $10^9$ bit sequences $x[n]$ and $y[n+k]$ as a function of the relative delay $k$. The solid line shows the theoretical median level one would expect for two independent, unbiased $10^9$ bit sequences.

Although the parallel binary sequences exhibit low bias and undetectable interchannel correlation, when each channel is subjected to more extensive statistical testing, we observed statistically detectable deviations from ideal behavior. We attribute these deviations to nonideal detector characteristics, including the temporal response of the photoreceiver electronics, which can produce small correlations between successive bits in each channel. As we previously demonstrated [9], the statistical properties of each channel can be significantly improved by constructing the XOR between each signal and a time-delayed copy of itself. Delays as short as 26 bits were found to be sufficient to produce sequences that consistently pass all of the National Institute of Standards and Technology (NIST) statistical tests for randomness [17]. While, in this demonstration, the delayed XOR process was computed offline, it could easily be implemented using real-time hardware. In Fig. 4, we show the results of the 188 NIST statistical tests applied to a $10^9$ bit record obtained from the XORed sequence $x[n] \oplus x[n-26]$. In order to pass each of the statistical tests, the composite $p$ value must exceed $10^{-4}$, and there may be no more than 19 failures out of 1000 trials. (The random excursion variant test may have no more than 13 failures out of 561 trials.) The XORed data set passes all of the NIST statistical tests, and we obtain similar results for the second parallel channel. We also constructed a 20 Gb/s sequence by sequentially interleaving the two binary sequences (after XOR processing). The interleaved sequence was also observed to pass all of the requisite statistical tests, which further confirms the independence of the sequences [18].

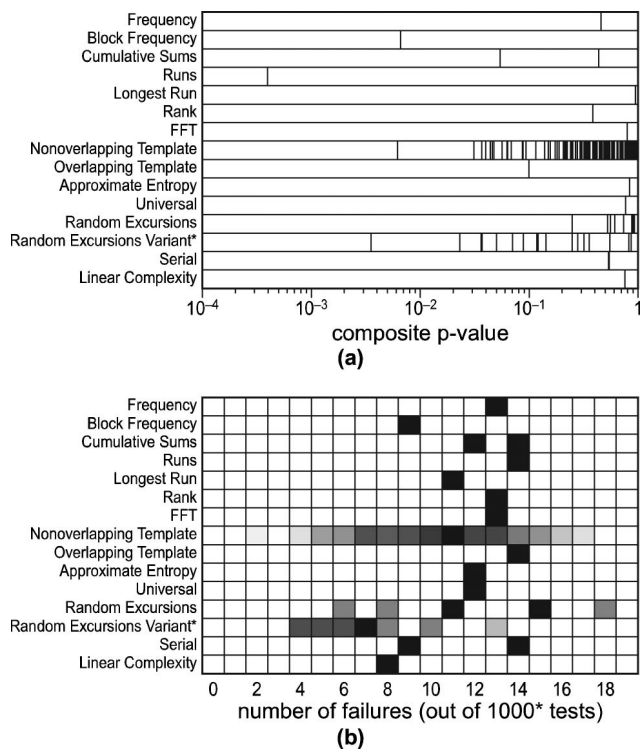

Fig. 4. Summary of the NIST test results. (a) Composite $p$ value obtained for each of the 188 statistical tests. (b) Gray-scale histogram reflecting the number of failures out of 1000 trials.

In summary, we report a scalable high-speed physical parallel random number generator based on ASE from a single optoelectronic light source. The system uses spectrally sliced light from a SLED to generate two parallel channels of random bits at 10 Gb/s each. With additional filters, the number of parallel wavelength channels could potentially be increased to at least 20, for a cumulative generation rate in excess of 200 Gb/s using a single SLED source. The system uses only commercially available optoelectronic components that could be integrated at the chip- or board-level for high-performance random number generation in computing applications.

The authors thank D. Donzis of Texas A&M University for useful insight about parallel random number generation and B. Ravoori, J. Salevan, and C. Williams for technical advice. X. Li acknowledges the support of the China Scholarship Council. This work is supported by a U.S. Department of Defense Multidisciplinary University Research Initiative grant (ONR N000140710734).